\documentclass[aps,twocolumn,groupedaddress,showpacs]{revtex4}
\usepackage{amsmath,amscd,amssymb,epsfig}
\begin{document}
\bibliographystyle{apsrev}

\title[]{Universal fluctuations in subdiffusive transport}

\author{I. M. Sokolov~$^1$, E. Heinsalu~$^{2,3}$,
P. H\"anggi~$^4$, and I. Goychuk~$^4$}
\affiliation{$^1~$Institut f\"ur Physik,
Humboldt-Universit\"at zu Berlin,
Newtonstra\ss e 15, D-12489 Berlin, Germany;
$^2$~National Institute of Chemical Physics and Biophysics, R\"avala
10, Tallinn 15042, Estonia; $^3$~IFISC, Instituto de F\'isica
Interdisciplinar y Sistemas
Complejos (CSIC-UIB), E-07122 Palma de Mallorca, Spain; $^4$~
Institut f\"ur Physik, Universit\"at Augsburg,
  Universit\"atsstr. 1,
  D-86135 Augsburg, Germany   }

\date{\today}

\begin{abstract}
Subdiffusive transport in tilted washboard potentials is studied
within the fractional Fokker-Planck equation approach, using the
associated continuous time random walk (CTRW) framework. The scaled
subvelocity is shown to obey  a universal law, assuming the form of
a stationary L\'evy-stable  distribution. The latter is defined by
the index of subdiffusion $\alpha$ and the mean subvelocity only,
but interestingly depends neither on the bias strength nor on the
specific form of the potential. These scaled, universal subvelocity
fluctuations emerge due to the weak ergodicity breaking and are
vanishing in the limit of normal diffusion.
The results of the analytical  heuristic theory are corroborated by
Monte Carlo simulations of the underlying CTRW.
\end{abstract}


\maketitle

A process of directed motion, for example the motion of a Brownian
particle under influence of a constant force, can be characterized by
its mean velocity $v$. The mean velocity is measured using a ruler and
a stopwatch in one of two  different setups: One can measure the
distances $L$ covered  within a fixed time interval $t$, or, like it
is done in sport competitions,  fix the distance $L$ and measure the
time intervals necessary to cover it. Thus, one can distinguish
between the the fixed time (FT) velocities and  the time-of-flight
(TOF) velocities. In any case, in ``normal'' situation  the typical
time necessary to overcome the distance $L$ grows on the average
linearly with $L$ in the TOF setup, or the typical distance covered
during the time $t$ grows on the average linearly with $t$ in the FT
setup. For our Brownian particle moving under the influence of
the constant force $F$ both setups give the values of $L/t$ which in
the limit of $t \rightarrow \infty$ for the FT measurement or $L
\rightarrow \infty$ for the TOF measurements reach the same sharp
value $v$.

For the case of biased anomalous diffusion (subdiffusion) the
situation drastically differs.
In what follows the subdiffusive motion $x(t)$
is modeled by  a continuous time random walk (CTRW)
with the
waiting time probability density (WTD) on sites following (for $t\gg\tau$)
a power law
\begin{equation}
\psi(t) \sim c(t/\tau)^{-1-\alpha}
\label{psi}
\end{equation}
with a diverging mean waiting time, i.e. with $0<\alpha <1$. In
(\ref{psi}), $\tau$ is a characteristic time scale and the prefactor
$c=[\tau \alpha\Gamma(1-\alpha)]^{-1}$ is introduced for simplicity
of further calculations.

For example,
charge transport processes in disordered, amorphous media
can be subdiffusive due to a trap-like transport mechanism with a
similar to (\ref{psi}) trapping time distribution
\cite{Scher,Hughes,Pfister,MetzlerPRL,HilferAnton,Allegrini}; 
an approximation which
can be justified for samples of macroscopic, but finite
size $L$ \cite{Pfister}. The corresponding
averaged current
$J(t)\propto d \langle \delta x(t)\rangle/dt$,
$\delta x(t)=x(t)-x(0)$, is a transient,
decaying to zero quantity \cite{Pfister,Hughes}.
However one can
define an {\it ensemble-}averaged mean subvelocity
$\overline{v}_{\alpha}=\Gamma(1+\alpha)
\langle \delta x(t)\rangle/t^{\alpha}$
 \cite{Goychuk06,Heinsalu06}
which is a quasi-stationary quantity for a
sufficiently large $L$ (neglecting finite size effects, i.e.
$L\to\infty$ when assuming
limit $t\to\infty$). In case of decaying photocurrent experiments
in thin amorphous films \cite{Pfister}, one can define an analogue of
subvelocity, namely
anomalous current
as $J_{\alpha}=\int_0^t J(t')dt'/t^{\alpha}=Q(t)/t^{\alpha}$,
where $Q(t)$ is the transferred charge. It  will first be quasi-stationary
and then decay anyway due to the finite size, edge effects.

The absence of a mean
trapping time leads to the (weak) ergodicity breaking
\cite{Bouchaud,BelBarkai} in the relevant transport processes.
In particular, the mean subvelocity of individual particles (before ensemble averaging)
is a random quantity as the time and ensemble averages
are not equivalent. This is just like the diffusion coefficient of individual particles
is a random quantity within our setup \cite{Lubelski,He}.
We shall consider a CTRW in a tilted periodic potential. In the continuous space
limit, it is described by the fractional Fokker-Planck equation
(FFPE) \cite{MetzlerPRL,Goychuk06,Heinsalu06}
\begin{eqnarray}\label{FFPE}
\frac{\partial^\alpha P(x,t)}{\partial t^\alpha}=
\kappa_{\alpha} \frac{\partial }{\partial
x} \left [ e^{-\beta U(x)} \frac{\partial}{\partial x} \, e^{\beta
U(x)} P(x,t)  \right ] \, ,
\end{eqnarray}
which we write down here in the form with the fractional Caputo derivative
$\partial^{\alpha}P(x,t)/\partial t^\alpha=(1/\Gamma(1-\alpha))
\int_0^{t}dt'[t-t']^{-\alpha}\partial P(x,t')/\partial t'$ 
\cite{HilferAnton,Goychuk06}.
In Eq. (\ref{FFPE}), $U(x)=V(x)-Fx$, where
$V(x+l)=V(x)$ is a periodic potential with period $l$, and $F>0$ is
the biasing force; $\beta=1/(k_BT)$ is
the inverse temperature, and $\kappa_{\alpha}$ is the subdiffusion coefficient.
The self-averaging in time does not occur \cite{Heinsalu06} and
mean subvelocity
remains a random variable even in the strict limit $t\to\infty$.
To find the corresponding probability distribution
presents the main objective of our Letter.
Only upon ensemble averaging, the
averaged value of subvelocity
coincides with one given by solving analytically the fractional
Fokker-Planck equation (FFPE) \cite{Goychuk06,Heinsalu06}.

We pose next the question about the universality class of
subvelocity distributions in this setup and find a simple result
which we confirm with numerical simulations. It is astounding that
this universality class does not depend on imposing a periodic
static potential $V(x)$ in addition to an applied constant force $F$
(what seems feasible experimentally). It also does not depend on
temperature given that  $\alpha$ is temperature-independent. This
universality feature is similar in nature with an established
universal scaling relation \cite{Goychuk06,Heinsalu06} between
anomalous current and biased diffusion,  originally suggested for
simple (i.e., in absence of a periodic potential) biased
subdiffusive CTRW transport \cite{Scher,Hughes}. It is derived below
by use of an heuristic argumentation, i.e. by reduction to a
coarse-grained CTRW. On the level of ensemble-averaged quantities,
we therefore obtain a universal law for the relative fluctuations of
(sub-)velocity, or fluctuations of anomalous current, which can be
tested experimentally.

{\it Theory for the biased CTRW.} We start out from a CTRW on a
one-dimensional lattice with period length $a$ and WTD (\ref{psi}).
The walk is biased and the nearest neighbors jumps (this assumption
is not restrictive and can be relaxed) occur with force-dependent
splitting probabilities $q^+$ (towards the  right) and $q^-$
(towards the left), with $q^{+}+q^{-}=1$. After $n$ steps, the mean
displacement is $L_n=\langle x\rangle=na(q^{+}-q^{-})$. From now on,
we measure distance $L$ in the units of $a^*(F)=a(q^{+}-q^{-})$.
Time will be measured in units of $\tau$ and the subvelocity
$v_{\alpha}$ in units of $v_0(F)=\Gamma(1+\alpha)a^*/\tau^{\alpha}$.

In the {\it fixed time  setup}, we fix the final time $t$ and ask
about the probability $p(n,t)$ to make $n$ steps. The answer is
well-known, see p. 248 in Ref. \cite{Hughes}: In the Laplace-domain,
it reads
\begin{eqnarray}
\hat p(n,u)= \frac{1-\hat\psi(u)}{u} [\hat\psi(u)]^n.
\end{eqnarray}
For $u\to 0$ (i.e. for $t\to\infty$), the leading term expansion of
the Laplace-transform of WTD in Eq. (\ref{psi}) is $\hat \psi(u)\sim
1-u^\alpha$. This leads to
\begin{eqnarray}\label{plu}
\hat p(n,u)\simeq u^{\alpha-1}\exp[n\ln(1-u^\alpha)]\simeq
u^{\alpha-1}\exp(-n u^{\alpha})
\end{eqnarray}
in the limit of large $n\to \infty$. The expression $\exp(-n
u^\alpha)$ is related to the characteristic function of the extreme
L\'evy stable law ${\cal L}_{\alpha}(t)$ of index $\alpha$ scaled with
parameter $n$, i.e. in the time domain it
corresponds to $n^{-1/\alpha}
\mathcal{L}_\alpha\left(n^{-1/\alpha}t\right)$. Considering $n$ as a
continuous parameter (distance $L=n$ in units of $a^*$) and noting
that Eq.(\ref{plu}) equals the Laplace transform of
\[
- \int_0^t \frac{d}{dL}\frac{1}{L^{1/\alpha}}
\mathcal{L}_\alpha\left(\frac{t}
{L^{1/\alpha}}\right)dt,
\]
we obtain, by a change of variable of
integration from $t$ to $\xi=t/L^{1/\alpha}$,
\[
p(L,t) \simeq - \frac{d}{dL} \int_{0}^{t/L^{1/\alpha}}
\mathcal{L}_\alpha(\xi) d\xi=
-\frac{d}{dL} C_\alpha \left( \frac{t}{L^{1/\alpha}} \right),
\]
where  $C_\alpha(x)$ is the cumulative distribution function of the
extreme L\'evy stable law, i.e.,
\[
p(L,t) \simeq \frac{1}{\alpha}\frac{t}{L^{1+1/\alpha}}
\mathcal{L}_\alpha \left( \frac{t}{L^{1/\alpha}} \right).
\]
Thus we can extract the distribution for the FT-subvelocity
$v_{\alpha} = \Gamma(1+\alpha)L/t^\alpha$ via a change of random
variable from $L$ to $v_{\alpha}$, yielding in terms of the scaled
subvelocity $\zeta_{\alpha}=v_{\alpha}/v_0(F)$, for all $F>0$, 
the stationary one-sided
L\'evy-stable distribution:
\begin{equation}
p(\zeta_{\alpha}) = \frac{\Gamma(1+\alpha)^{1/\alpha}}{\alpha
\zeta_{\alpha}^{1+1/\alpha}} \mathcal{L}_\alpha\left
[\left(\frac{\Gamma(1+\alpha)}{\zeta_{\alpha}}\right)^ {1/\alpha}
\right] \;.
\label{velocity}
\end{equation}
This universal form of the subvelocity distribution presents a
major result of our study.

Let us demonstrate that the very same result is recovered also
within the {\it time-of-flight  setup}.
That is, we are looking for the asymptotic distribution of times to
make a large number of $n$ steps. The corresponding distance  will
assume a sharply peaked distribution around its mean value which can
be identified with the sample size $L$. The random time $t$
necessary to traverse the system of length $L$ in the TOF setup is
essentially the time necessary to make $n$ steps. The overall time
to make $n \gg 1$ steps tends in distribution to a one-sided L\'evy
law $n^{-1/\alpha}\mathcal{L}_{\alpha}\left(n^{-1/\alpha}t\right)$.
To see this it is sufficient to notice that the Laplace transform of
the probability to find this time  is given by $\hat
p(u)=[\hat\psi(u)]^n \simeq (1-u^\alpha)^n \simeq \exp[n
\ln(1-u^\alpha)] \simeq \exp(-nu^\alpha)$. The distribution of
$v_\alpha$ is then be obtained by the same change of variable as
used above (the only difference being that $t$ is now a random
variable and $L$ is fixed) to yield the
 same result in (\ref{velocity}).

The averaged value of the scaled subvelocity with
distribution (\ref{velocity}) is
one, $\overline{\zeta}_{\alpha}=1$, i.e. the subvelocity in (\ref{velocity}) is
scaled in fact through its averaged value $\overline{v}_{\alpha}(F)=v_0(F)$.
All the higher moments can be obtained using the change
of variable $y=[\Gamma(1+\alpha)/v_{\alpha}]^{1/\alpha}$ and
the relation
\begin{equation}
\int_0^\infty y^\eta \mathcal{L}_\alpha(y)dy = \frac{\Gamma(1-\eta/\alpha)}
{\Gamma(1-\eta)}
\label{Sato}
\end{equation}
which is valid for any $\eta \in (-\infty,\alpha)$, see in Ref.
\cite{Sato}. With $\eta = -2\alpha$, it yields the second moment
\begin{eqnarray}
\overline{\zeta_{\alpha}^2}=\frac{2\Gamma^2(1+\alpha)}{\Gamma(1+2\alpha)}\;.
\end{eqnarray}
The relative fluctuation of subvelocity $\delta
v_{\alpha}=\sqrt{\overline{v^2_{\alpha}}-(\overline{v}_{\alpha})^2}/
\overline{v}_{\alpha}$ equals  in fact the universal scaling
relation between the averaged subdiffusion current and biased
diffusion of Refs. \cite{Scher,Hughes,Goychuk06,Heinsalu06}, i.e.,
\begin{eqnarray}\label{fluctuation}
\delta v_{\alpha} & = &
\frac{\sqrt{\overline{\delta x^2(t)}}}{\overline{\delta x(t)}}
=\sqrt{\frac{2\Gamma^2(1+\alpha)}{\Gamma(1+2\alpha)}-1}\\
& = &\lim_{t\to\infty}
\frac{\sqrt{\langle \delta x^2(t)\rangle}}{\langle \delta x(t)\rangle }\;.
\nonumber
\end{eqnarray}
This result is not trivial: This is so because $\overline{(...)}$ is
the average over the stationary subvelocity distribution
$p(v_{\alpha})=p(\zeta_{\alpha}=v_{\alpha}/\overline{v}_{\alpha})/
\overline{v}_{\alpha}$, while $\langle .... \rangle$ is the average
over the time-dependent population probabilities $p_i(t)$ of the
lattice sites. This constitutes our second main result; it shows
that weak ergodicity breaking is at the root of this  remarkable
scaling relation (\ref{fluctuation}). Namely, it is responsible for
the startling change of the law of subdiffusion from $\langle \delta
x^2(t)\rangle \propto t^\alpha$ when $F=0$  to $\langle \delta
x^2(t)\rangle \propto t^{2\alpha}$ at $F\neq 0$, i.e. subdiffusion
turns over into superdiffusion for $0.5<\alpha<1$. 
We remark, however, that the scaling relation (\ref{fluctuation})
between the current and the biased diffusion
cannot be employed to deduce the main result in (\ref{velocity}).

\begin{figure}[t]
\centering
\includegraphics[width=7.5cm]{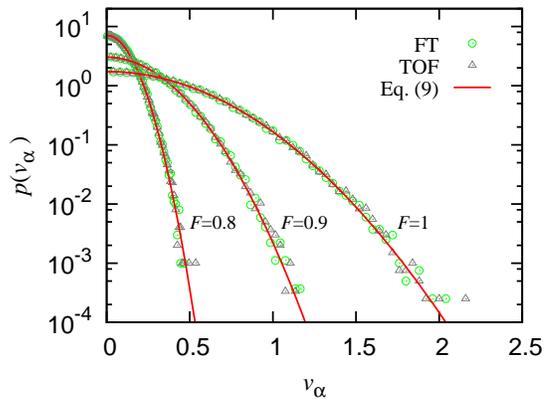}
\caption{(Color online) Numerical subvelocity distribution
$p(v_{\alpha})$ for $\alpha=0.5$ in both FT and TOF setups for a
periodic potential $V(x)=V_0\cos(2\pi x/l)$ and differing bias
forces $F$. $v_{\alpha}$ is scaled in units of $v_0^*(F)=
\overline{v}_{\alpha}(F)/v_{cr}$, where $\overline{v}_{\alpha}$ is
given by  (\ref{stratSUB}) and
$v_{cr}=F_{cr}\kappa_{\alpha}/(k_BT)$. The solid lines depict the
theoretical result (\ref{one-half}):
$p(v_{\alpha})=p(\zeta_{\alpha}=v_{\alpha}/v_0^*)/v_0^*$.
}
\label{Fig1}
\end{figure}

In particular, for $\alpha=0.5$ Eq. (\ref{velocity}) simplifies
to one-sided Gaussian form (cf. Fig. \ref{Fig1})
\begin{eqnarray}\label{one-half}
p(\zeta_{1/2})=\frac{2}{\pi}\exp\left(- \frac{1}{\pi}
\zeta_{1/2}^2\right),
\end{eqnarray}
and $\delta v_{1/2}=\sqrt{\pi/2-1}$. For other values of $\alpha$, a
handy approximation to the subvelocity distribution can be obtained
using the corresponding small-$x$ asymptotic behavior of the
Levy-stable distribution \cite{Sato,SokolovBelik}. It reads,
\begin{eqnarray}\label{approx}
p(\zeta_{\alpha})\simeq A(\alpha) (\alpha
\zeta_{\alpha})^{\frac{\alpha-1/2}{1-\alpha}}\exp[-B(\alpha)
\zeta_{\alpha}^{1/(1-\alpha)}],
\end{eqnarray}
where
$A(\alpha)=[\sqrt{2\pi(1-\alpha)}\Gamma(1+\alpha)]^{-1}$
and $B(\alpha)=(1-\alpha)\alpha^{\alpha/(1-\alpha)}
\Gamma(1+\alpha)^{-1/(1-\alpha)}$.
For $\alpha=0.5$ this approximation reproduces the exact result in
 (\ref{one-half}). For $0.5<\alpha<1$, it correctly predicts that
the distribution function is non-monotonic, possessing a maximum,
cf. Fig. \ref{Fig2} below, which becomes sharp for $\alpha\to
1$. In this limit, 
the relative fluctuation vanishes, $\delta v_{\alpha}\to 0$, and
the velocity distribution tends to the delta-function centered at
$\overline{v}_{\alpha}$.
However, the correct value $p(0)$ always remains
finite for $\alpha<1$, implying that there are always particles
which become immobilized. For $\alpha \leq 0.5$, $p(v_{\alpha})$
decays monotonically. Moreover, for small $\alpha\to 0$, the
distribution becomes nearly exponential, consistent with $\delta
v_{\alpha}\to 1$ in this limit, see Fig. \ref{Fig3} below.
%

All our analytical findings are confirmed by the numerical
simulations  of the underlying CTRW in a biased cosine potential
$V(x)=V_0\cos(2\pi x/l)$, using the Mittag-Leffler distribution
$\psi(\tau)$ and the numerical algorithm detailed in
\cite{Heinsalu06,remark}. It is surprising that all these results
hold {\it universally}, i.e. these are independent of the details of
periodic potential and the temperature. To elucidate this astounding
fact, being numerically confirmed with Figs. \ref{Fig1}, \ref{Fig2},
\ref{Fig3} for a washboard potential (details are given below), we
make use of the reasoning put forward with  Ref. \cite{Goychuk06}.

\begin{figure}[t]
\centering
\includegraphics[width=7.5cm]{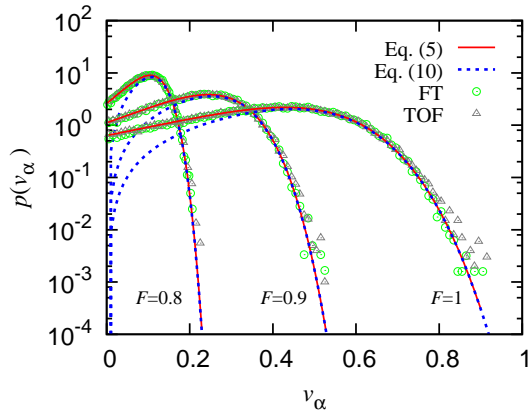}
\caption{(Color online) Numerical subvelocity distribution
$p(v_{\alpha})$ {\it vs.} the  analytical approximation in Eq.
(\ref{approx}) and the exact result in Eq.  (\ref{velocity}) for
$\alpha=0.8$. The same scaling applies as in  Fig. \ref{Fig1}. }
\label{Fig2}
\end{figure}

\begin{figure}[t]
\centering
\includegraphics[width=7.5cm]{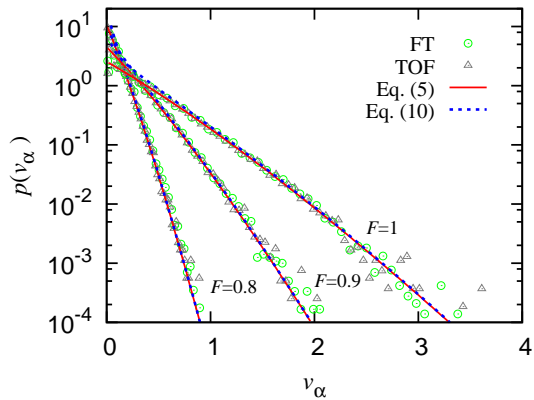}
\caption{(Color online) Same as in Fig. \ref{Fig1} for $\alpha=0.2$.
 }
\label{Fig3}
\end{figure}

{\it Theory for  washboard potentials.} We dilate the lattice by
introducing many more points with separation $\Delta x\to 0$. The
residence time distributions on every point are chosen to be
Mittag-Leffler distributions \cite{HilferAnton} with different time
scaling parameters $\tau_i=1/\nu_i$. This distribution belongs to
the same class in  (\ref{psi}). Each point $i$ is characterized also
by the  right and left nearest neighbor jump probabilities
$q^{\pm}_i=g_i^{\pm}/(g_i^{+}+ g_i^{-})$, and by the fractional
forward and backward rates, $g_i^{\pm}=q_i^{\pm}\nu_i^{\alpha}$,
respectively. These quantities follow from the potential $U(x)$ as
so that the Boltzmann relation $g_{i-1}^{+}/g_i^{-}=
\exp[\beta(U_{i-1}-U_i)]$ is fulfilled. Here, $U_i \equiv U(i \Delta
x)$, $U_{i \pm 1/2} \equiv U(i \Delta x \pm \Delta x /2)$, and
$\nu_i=(g_i^{+}+g_i^{-})^{1/\alpha}$. The generalized master
equation for such a CTRW is \cite{HilferAnton,Goychuk06}:
\begin{eqnarray*}
\frac{\partial^{\alpha} P_i(t)}{\partial t^\alpha}
\!\! = \!\! g_{i-1}^{+} \, P_{i-1}(t) + g_{i+1}^{-}
\, P_{i+1}(t) - (g_i^{+} + g_i^{-}) \, P_i(t)\;.
\end{eqnarray*}
In the spatial continuous limit $\Delta x\to 0$, it yields the FFPE
(\ref{FFPE}). In this way, we simulate the stochastic dynamics
associated with (\ref{FFPE}) on a sufficiently dense grid with step
$\Delta x$, using the Monte Carlo algorithm  from \cite{Heinsalu06}.

Consider next a periodic potential with period $l$ subjected to a
finite bias force $F$. One can course-grain the corresponding
limiting CTRW and to map it onto a new biased CTRW with the lattice
period $l$, i.e. we average over spatial period $l$. The precise
form of the coarse-grained WTD is not known. However, it belongs to
the same class as  (\ref{psi}); only the  time parameter $\tau$ is
correspondingly changed together with the coarse-grained splitting
probabilities $q^{\pm}$.
We note that such course-graining of Markovian, normal diffusion in
washboard potentials yields a {\it non-Markovian}  CTRW which can
give rise to such profound effects as giant acceleration of
diffusion \cite{giant}. In clear contrast, our original CTRW is
already a profoundly non-Markovian, non-ergodic process possessing
infinite memory. Coarse-graining it further does not change the
universality class because
no correlations between the residence times in non-overlapping
spatial domains occur.

For  arbitrary periodic tilted potentials, the result for
ensemble-averaged subvelocity was obtained  in Refs.
\cite{Goychuk06,Heinsalu06} as:
\begin{eqnarray}\label{stratSUB}
\overline{v}_{\alpha}(F)  = \! \! \frac{ \kappa_{\alpha} l
\, [1 - \exp(-\beta
F l)]}{\int_{0}^l \mathrm{d} x \int_{x}^{x+l} \mathrm{d} y \,
\exp(-\beta[U(x) - U(y)])} \, .
\end{eqnarray}

In all our numerical simulations we used the archetype cosine
potential $V(x)=V_0\cos(2\pi x/l)$. The grid contains $200$ points
within each spatial period. A scaled temperature of $k_B T=
0.1\;V_0$ is used throughout and the force $F$ is scaled in units of
the critical force $F_{cr}$, where with $F>F_{cr}$  
the potential $U(x)$ becomes
monotonic without barriers in between. The number of particles is
$N=10^5$. The different lines for fixed $\alpha$ and different bias
values $F$, are due to  the different values of the scaling
parameter $\overline{v}_{\alpha}(F)$, calculated in accordance with
(\ref{stratSUB}). In accordance with our theory, all the related lines
perfectly  coincide (not shown) after re-scaling $v_{\alpha}\to
\zeta_{\alpha}=v_{\alpha}/\overline{v}_{\alpha}(F)$, $p \to p\cdot
\overline{v}_{\alpha}(F)$, for all $F>0$. 
The numerical results thus corroborate
with theory.

In conclusion, we have shown that the weak ergodicity breaking
is responsible for the universal scaling relation
(\ref{fluctuation}) between the anomalous current and subdiffusion
occurring in arbitrary tilted periodic potentials.  This intriguing
result follows from the universal law for the theoretically
deduced subvelocity distribution  in (\ref{velocity})
which is the major finding of this work.

The authors  acknowledge the support by the DFG within SFB 555 (I.
Sokolov), by the Estonian Science Foundation through Grant  No. 7466
and Spanish MEC and FEDER through project FISICOS (FIS2007-60327, E.
Heinsalu), and by the  German Excellence Initiative via the
Nanosystems Initiative Munich (NIM, P. H\"anggi, I. Goychuk).


\begin{references}
\bibitem{remark} In this work, we use the exact generator for Mittag-Leffler
distribution from Ref. \cite{Kozubowski}.
\end{references}
\end{document}